\documentclass[12pt]{iopart}
\usepackage[utf8]{inputenc}
\usepackage[T1]{fontenc}

\usepackage{array}
\usepackage{booktabs}

\usepackage{multirow}

\usepackage{graphicx}

\usepackage[square,comma, sort&compress,numbers]{natbib}
\usepackage{doi}
\usepackage{hyperref}

\renewcommand\textasciitilde{{\lower.8ex\hbox{$\mathtt{\char`\~}$}}}

\begin{document} 
\title[The N400 for Brain Computer Interfacing]{The N400 for Brain Computer Interfacing: complexities and opportunities}

\author{K V Dijkstra, J D R Farquhar and P W M Desain}
\address{Radboud University, Donders Institute for Brain, Cognition and Behaviour }

\begin{abstract}

The N400 is an Event Related Potential that is evoked in response to conceptually meaningful stimuli. It is for instance more negative in response to incongruent than congruent words in a sentence, and more negative for unrelated than related words following a prime word. 
This sensitivity to semantic content of a stimulus in relation to the mental context of an individual makes it a signal of interest for Brain Computer Interfaces. Given this potential it is notable that the BCI literature exploiting the N400 is limited. We identify three existing application areas: (1) exploiting the semantic processing of faces to enhance matrix speller performance, (2) detecting language processing in patients with Disorders of Consciousness, and (3) using semantic stimuli to probe what is on a user's mind. 
Drawing on studies from these application areas, we illustrate that the N400 can successfully be exploited for BCI purposes, but that the signal-to-noise ratio is a limiting factor, with signal strength also varying strongly across subjects. Furthermore, we put findings in context of the general N400 literature, noting open questions and identifying opportunities for further research.

\end{abstract}

\maketitle

\section{Introduction}

A Brain Computer Interface measures brain activity to infer something about the user and take appropriate action based on that inference. This can be the passive recognition of the user's mental state, or the detection of an intention that the user is actively trying to transmit. 
In an ideal world, the user's intention would be extracted directly from their brain activity: for instance, if they wish to convey a message, the user simply thinks of the words and the system decodes the brain signature(s) that accompany them. In practice this is generally not possible, as the representation of such thoughts is currently unknown and the limited temporal or spatial resolution of current neuroimaging systems likely precludes them from real-time detection. Advancements have been made in decoding or reconstructing speech from invasive electrophysiological recordings (intracranial or intracortical), with the aim to extend these to imagined speech (see Martin et al. \cite{martin_use_2016} for a recent overview), but this field is still young and the invasive nature limits wide applicability.

In absence of the ability to measure a desired action directly from someone's brain activity, it can be useful to look at brain signatures or signals that we know \textit{can} be measured reliably from brain activity, and center the design of a Brain Computer Interface around those. For instance, using electroencephalography (EEG), Event Related Potentials (ERPs) can be measured that reflect the brain activity in response to stimuli. The P300 is an ERP that is elicited for stimuli that are relevant to the task a user is performing in a stream of non-target stimuli \cite{pritchard_psychophysiology_1981}. By designing a stimulus set such that the subject has the option to determine which stimuli are task-relevant (by directing their attention to the stimuli associated with the selection of interest), a P300 BCI can enable a user to make multiple-choice selections. A well known example is the P300 speller. When flashing the columns and rows of the letter matrix, the user's attention to a single letter elicits P300s only for flashes on that letter's row/column coordinates. This subsequently enables the system to infer (the position of) the attended letter based on the extracted ERP responses. Other examples of such an approach are using neurofeedback of the mismatch negativity for sound discrimination learning \cite{chang_unconscious_2014}, or using the power changes in the posterior alpha associated with (covert) spatial attention for indicating a direction \cite{van_gerven_attention_2009,treder_brain-computer_2011}.

The N400 is another well established ERP in neuroscience literature. It was first discovered when in sentential contexts, incongruent sentence endings exhibited a more negative ERP than congruent sentence endings, peaking at around 400 ms \cite{kutas_reading_1980}. In the decades since, it has been investigated extensively (see Kutas \& Federmeier \cite{kutas_thirty_2011} for an overview). The N400 is not limited to sentence contexts, and can also be elicited through semantic priming: presenting a prime word followed by a related or unrelated stimulus, with the unrelated stimuli eliciting a more negative N400 \cite{bentin_event-related_1985}. This extends to cases where the prime is not presented, but merely actively recollected \cite{van_vliet_guessing_2010}. Broadly speaking, it is sensitive to the relation of the presented stimulus to the mental context of the user. 

This is interesting from a BCI perspective, because it suggests that the N400 could be used to infer information about the user's mental context, without the user having to make this explicit. 
That is, given that BCIs are (currently) unable to read what someone is thinking directly from their brain activity, the N400 could potentially be used to infer information about the user's active mental context by presenting stimuli with specific content, and observing the size of the elicited N400 to these stimuli.
However, in addition to these contextual effects, numerous (inherent) stimulus characteristics have been shown to also affect the N400: for instance, simply the frequency that a word occurs in a language can affect the N400 amplitude. These may need to be controlled or accounted for when using the N400 for a BCI application.

Given the prominence of the N400 in the neuroscience literature, attention for the N400 in the BCI field has been comparatively limited. This may reflect a disinterest in the kind of BCI paradigms that could be designed around a N400 effect, or, a (perceived) difficulty to make such paradigms work. To obtain more insight, we searched for the keywords "Brain Computer Interface" and "N400" and identified three research lines that do make use of an N400 effect: (1) the use of the N400 in response to familiar face stimuli for boosting performance of the matrix speller \cite{kaufmann_flashing_2011,kaufmann_face_2013,jin_erp-based_2014}, (2) to detect the absence or presence of language processing in patients with impaired levels of consciousness (Disorders of Consciousness) \cite{kotchoubey_information_2005,steppacher_n400_2013,daltrozzo_cortical_2009}, and (3) to probe for information about the active mental state of the user, e.g., a category or word on the user's mind, by presenting stimuli that elicit responses of relatedness with regard to this mental context \cite{wenzel_real-time_2017,geuze_detecting_2013,dijkstra_semantic_2018}. 

In the following sections we discuss these research lines to provide an overview of the current ways the N400 is exploited for BCI purposes, to determine limitations, and to identify opportunities for improving existing paradigms or developing new paradigms. To provide context, we start with a more detailed description of the N400 and the conditions in which it is elicited. The Neuroscience literature on this N400 effect is extensive, and a good overview article already exists \cite{kutas_thirty_2011}, so the focus here is not to be comprehensive, but to summarise the most important aspects and to take the opportunity to discuss its properties with a BCI context in mind.
With this N400 overview and an analysis of the limitations and potential opportunities in existing N400 BCI research, we hope to provide a starting point for researchers with an interest in using the N400 for a BCI paradigm, in deciding whether or not their intended application may be feasible. 

\section{Overview of the N400}

\begin{figure}
\centering
\includegraphics{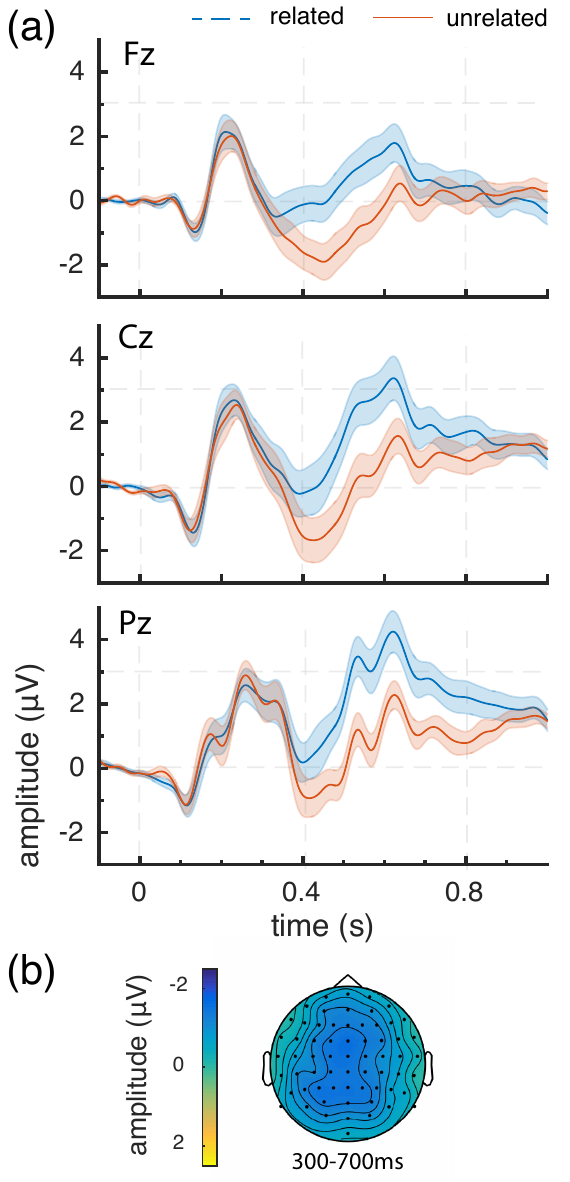}
\caption{Brain responses to related and unrelated stimuli-words in relation to a memorised target-word (data from \cite{dijkstra_semantic_2018}). (A) Grand average ERPs for channels Fz, Cz and Pz, for 20 subjects. Shaded areas represent a bootstrapped 95\% confidence interval of the mean (B) grand average topography for the 300-700 ms period, (unrelated - related) }
\label{fig:N400_example}
\end{figure}

The N400 is characterised as a negative going wave that reaches its maximum (negative) amplitude around 400ms after stimulus presentation over centro-parietal electrodes \cite{kutas_thirty_2011}. It is most known for its sensitivity to semantic context, being more negative to incongruent than congruent words in a sentence \cite{kutas_reading_1980}, or more negative for unrelated than related words following a prime word \cite{bentin_event-related_1985}. Figure~\ref{fig:N400_example} depicts the N400 in response to related and unrelated stimulus words in relation to a remembered target word, showing the characteristic negativity, latency and topography \cite{dijkstra_semantic_2018}. One experiment comparing the word pair priming and sentence congruency effects found a N400 difference of 3.4 $\mu V$ (Cz) and 5.2 $\mu V$ (Pz) for sentences, and a difference of 1.8 $\mu V$ (Cz) and 1.9 $\mu V$ (Pz), for word pair stimuli, suggesting that the sentence congruency effect is stronger than the word-priming effect.

Repetition of a stimulus within an experimental session can also affect the N400 to that stimulus, with subsequent presentations having a reduced N400 amplitude \cite{rugg_event-related_1989,rugg_event-related_1990}. This reduction is larger the sooner the repetition occurs, but persists even when e.g., 19 intervening words are presented before the item is repeated, in a list \cite{rugg_event-related_1989}, or when up to several hundred words are read before encountering the repetition in a text \cite{van_petten_interactions_1990}. A 15 min break, on the other hand, can be sufficient to make the repetition effect nearly undetectable \cite{rugg_event-related_1990}. 

In addition to these contextual manipulations, the N400 amplitude has been shown to be sensitive to certain inherent stimulus properties, that can be measured even in isolated stimulus presentations. Instances of properties that affect the N400 amplitude for word stimuli are: the (written) frequency with which a word occurs in a language (larger N400 for low frequency words) \cite{rugg_event-related_1990,petten_comparison_1993}, the concreteness of a word (larger N400 for concrete than abstract words) \cite{barber_concreteness_2013,kounios_concreteness_1994}, and the orthographic neighbourhood size (i.e., how many words are similar in spelling; a larger N400 for words with a larger/denser neighbourhood) \cite{laszlo_n400_2011, holcomb_electrophysiological_2002}. The latter is only one of several neighbourhood effects, where a neighbourhood can be defined both orthographically and associatively (which words are related to this word). In both definitions a larger neighbourhood size, and even a higher word frequency of the neighbours can increase the N400 amplitude \cite{laszlo_n400_2011}. 

Experiments where both contextual and lexical effects have been manipulated, show that these effects can interact: for instance, both stimulus repetition and higher word frequency lead to a smaller N400 amplitude in isolation, however, when both low and high frequency words were presented twice, the effect of the repetition on the high frequency word's ERP was small to none, while the repetition of the low frequency words reduced the size of the N400 \cite{rugg_event-related_1990}. Kounios et al. find a similar interaction between concreteness and repetition \cite{kounios_concreteness_1994}. A study by Dambacher et al. shows another interaction, that of sentence context (measured as word predictability), and word frequency, with the effect of word frequency decreasing as predictability increases \cite{dambacher_frequency_2006}.

Another BCI relevant interaction effect is that of age, where for certain experimental manipulations the N400 amplitude decreases with age (reminiscent of the changes in the P300 with age \cite{van_dinteren_p300_2014}), but does not decrease for others. For instance, while sentential context effects have been shown to elicit smaller N400 differences in older subjects \cite{federmeier_sounds_2003,payne_contextual_2018}, Payne et al. \cite{payne_contextual_2018} found that this did not extend to the lexical effects they tested (i.e., word frequency and orthographical neighbourhood). Federmeier et al. \cite{federmeier_sounds_2003} found a reduction of the N400 for sentence final-word (in)congruency, for their older subjects, but no reduced N400 difference for lexical association (i.e., priming) effects, within those sentences. 

Returning to the semantic context effects, research indicates that it is not only the `local' (e.g., sentence level) semantic context in which a stimulus occurs that is relevant for the N400 amplitude. Discourse (i.e., context on a larger scale), built up over a few preceding sentences, can increase the N400 amplitude when violating expectations from previous context \cite{van_berkum_anticipating_2005}, or decrease the N400 amplitude when prior context causes a normally incongruent sentence ending to be interpretable as congruent \cite{nieuwland_when_2006}. Furthermore, beyond active mental context, even passive world-knowledge can elicit a larger negativity when it is incongruent with a presented stimulus: world-knowledge violations, e.g., a sentence asserting that that dutch trains are \textit{white}, elicited a negativity similar to the N400 from semantic violations, e.g., a sentences asserting dutch trains are \textit{sour} \cite{hagoort_integration_2004,hagoort_semantic_2009}. This example, however, does not distinguish between the kind of world knowledge that stems from a learned association between concepts (e.g., `train' and `yellow'), and world knowledge that allows truth-values to be assigned to statements (propositional knowledge). In the latter case the relation with the N400 may not always be straightforward (see e.g., the effect of negation on the N400 \cite{fischler_brain_1983,nieuwland_when_2008}).

\begin{table}[p!]
\centering
\vspace*{-0.2in}
\footnotesize
\caption{
{\label{tab:example}Summary of the main variables affecting the N400 amplitude, with a short description. The direction of the effect is noted, with '$>$' indicating which will obtain a larger N400 amplitude (i.e., have a more negative ERP in the N400 timerange). }
}
\begin{tabular}{m{1.7cm}m{3.2cm}m{6cm}m{4.5cm}}
\br
Level & Variable & Description & Direction  \\
\mr

\multirow{4}[50]{1.3cm}{\centering lexical effects} & word frequency & How often a word occurs in (written) language. \textit{House} is a frequent word, \textit{parsimonious} is not. & low freq. $>$ high freq. \\ \addlinespace

& word neighbourhood & Orthographic neighbourhood: words that are similar in spelling to the word in question (visual presentation) \textit{Mine} has neighbours such as \textit{mane}, \textit{line}. \newline
Phonological neighbourhood: words that are similar in sound  (auditory presentation) \newline
The neighbourhood size (or density) refers to the number of words that are within close distance of a given word.  & large  $>$ small neighbourhood \newline dense  $>$ sparse neighb. \\ \addlinespace

& word concreteness & Concreteness  (as opposed to abstractness) describes the degree with which a word can be experienced with the senses\cite{brysbaert_concreteness_2014}). \textit{House} is a concrete word, \textit{justice} is abstract. 
 & concrete $>$ abstract \\ 
 \mr

\multirow{11}[50]{1.3cm}{\centering contextual effects} & stimulus repetition & Whether it concerns the first presentation of a stimulus or a second or n\textsuperscript{th} presentation. The interval (time before repetition) is also relevant. & first presentation $>$ repetition \newline long interval $>$ short interval \\ \addlinespace

& lexical relatedness or association  & Semantic priming occurs when a word was preceded by a semantically similar word (\textit{cup - bowl}) or associated word (\textit{tea - cup}) compared to an unrelated or unassociated word. & unrelated $>$ related \newline  unassociated $>$ associated \\ \addlinespace

& sentence context & How well does a word fit in the existing sentence semantic context. Does it fit with the sentence so far (:=congruent) or is it anomalous (:=incongruent). \newline
How often would people end the sentence word the word in question (:= cloze probability). & incongruent $>$ congruent \newline  low cloze-prob. $>$ high prob. \\ \addlinespace

& discourse context or world-knowledge & How well does a word fit in the context of a larger preceding text, or with existing world-knowledge & incongruent $>$ congruent \\ \addlinespace

& non linguistic context & Other types of meaningful stimuli in the form of e.g., drawings, photos, videos, sounds and smells representing concepts such as objects, actions, faces and even mathematics \cite{kutas_thirty_2011} can elicit N400 effects & incongruent $>$ congruent \newline un-primed $>$ primed \\

\br
\end{tabular}

\normalsize
\end{table}

An overview of the mentioned effects can be found in table~\ref{tab:example}. Here the different variables that have an effect on the N400 are listed, with a short description and the direction of the effect. 

It is important to note that whilst most N400 studies have focused on visually presented word stimuli, the differential effect with respect to the active mental context has also been demonstrated with other presentation modalities. Spoken words elicit N400 effects analogous to the visual presented stimuli \cite{holcomb_auditory_1990,winsler_electrophysiological_2018} (with phonological neighbourhood as the auditory analogue of orthographic neighbourhood). Furthermore, they are not limited to words: pictures of faces, for instance, have been found to elicit N400 (priming) effects \cite{barrett_event-related_1989,eimer_event-related_2000,bentin_effects_1994}. More broadly, actions or objects, represented with pictures (drawings or photo's), videos, sounds and even smells have successfully been used to elicit N400 effects (see Kutas et al. (2011) \cite{kutas_thirty_2011}). This does not mean that the N400 effects are identical across modalities. For instance, in a priming experiment Holcomb \& Neville found a larger N400 amplitude and an earlier N400 onset for auditory stimuli than for the same stimuli presented visually \cite{holcomb_auditory_1990}.

From a BCI standpoint it is also relevant to know whether subjects need an explicit task to elicit the signal. For the N400 it is generally sufficient for subjects to attend the stimuli to elicit the effects, as processing stimuli for meaning happens automatically. This is, unless the task explicitly does not require people to process the stimuli as a stimulus containing meaning at all (e.g., just the size of the stimulus on the screen), in such a case the N400 may be absent \cite{chwilla_n400_1995,deacon_relationship_1991}. Additional experimental manipulations have shown that the processing of the stimuli does not need to reach conscious awareness for a detectable N400, with both attentional blink and masked priming studies finding effects of priming on the N400 (see Kutas \& Federmeier \cite{kutas_thirty_2011} and Deacon \& Shelley-Tremblay \cite{deacon_how_2000} for overviews). However, the degree of attention does play a role in the strength of the elicited N400 effect. Cruse et al. \cite{cruse_reliability_2014}, for instance, link the degree of active semantic task involvement of the subjects to the number of subjects for which a N400 can be detected, finding progressively smaller N400s for a covert (mental decision only) and passive task (no task), compared to an overt task (requiring a behavioural response).

The functional explanation of the N400 is still a matter of debate. To summarise the main views, the N400 has been suggested to reflect brain activity from the activation of (the representation of) the stimulus in long term memory, to reflect activity of the integration of the stimulus in the active mental context, or, to reflect a combination of such processes \cite{lau_cortical_2008,hagoort_semantic_2009,kutas_thirty_2011,baggio_balance_2011,nieuwland_dissociable_2019}. In the first view the amplitude of the N400 is reduced when the stimulus' representation was (partially) pre-activated by prior stimuli, facilitating semantic access. In the integration account the N400 amplitude reflects the difficulty with which the stimulus can be incorporated into the previously accrued context.
There have also been attempts at building a computational model of the N400: a recent paper models the N400 as reflecting the degree of change in probabilities in an update of the mental model in response to the presentation of a given stimulus. This computational model subsequently exhibits a considerable number of the N400 effects outlined above \cite{rabovsky_modelling_2018}.

In addition to the N400, language manipulations sometimes elicit a later positive ERP: the late positive component or complex (LPC), or P600 \cite{van_petten_prediction_2012,osterhout_event-related_1992,leckey_p3b_nodate,kos_individual_2012}. It can co-occur with the N400 \cite{van_petten_prediction_2012}, or be elicited independently, e.g., in response to syntactic violations \cite{osterhout_event-related_1992}. In sentence congruence paradigms, it does not occur as consistently as the N400, with various studies not finding a positivity in this late range (600-900 ms) \cite{van_petten_prediction_2012}. Some studies attempt to interpret this late response as a late P300 (P3b) \cite{sassenhagen_p600-as-p3_2014}. In fact, it is unclear whether these late effects represent a single brain response, or multiple functionally dissociable responses \cite{leckey_p3b_nodate}. Furthermore, in a study by Kos et al. \cite{kos_individual_2012} only about half the participants (n = 72) exhibited a LPC, while others showed an extended negativity of the N400. This may cause the effect to be averaged out in grand average ERPs, while there may yet be a signal present for individual subjects.

In summary, the N400 is a complex response that has been shown sensitive to many parameters of the user or user's mental context, ranging from general-world-knowledge and subject age, to stimulus properties (such as word frequency, orthographic or phonological neighbourhood size and concreteness), from long range document-context effects, to short range sentence congruency effects and direct prime-probe relations or repetition effects. These multiple levels of interaction make the N400 response both potentially attractive for many possible BCI applications, and particularly challenging for these possible applications to control for all the other possible interactions. A similar complexity (though much less extreme) is seen in p300 spellers where choices with regard to the matrix size affect the size of the elicited P300 \cite{allison_erps_2003}, which may be attributable to the fact that the P300 is not only sensitive to the task relevance of a stimulus (i.e., is it a 'target' for the current selection), but also the inherent stimulus frequency or probability, with lower probability stimuli eliciting larger P300s. 

\section{Current approaches using N400 for BCI purposes}
We have identified three categories of existing research that use the N400 for BCI purposes: (1) approaches that exploit the N400 to enhance existing BCI applications, (2) approaches that aim to detect language processing in patients with Disorders of Consciousness (DoC), (3) and approaches that use stimuli to probe the users mental context. We discuss these here, identifying limitations, open questions and opportunities for further research.

\subsection{Enhancing existing BCI applications}
We mentioned briefly in the previous section that the N400 can also be elicited in response to (pictures of) faces. This is both in the priming sense \cite{barrett_event-related_1989}, effects of repetition \cite{bentin_effects_1994}, and by presenting faces outside of context, where, similar to the lexical effects in language, differences have been identified in response to familiar and unknown faces \cite{eimer_event-related_2000}. Specifically, familiar faces elicit a larger N400 than the unknown faces, which appears at odds with the frequency effects in words, where pseudo-words elicit larger N400s than real words. A suggested reason for this is that the unknown faces do not initiate semantic processing and show no N400 negativity at all \cite{bentin_structural_2000}. This used to be similarly suggested for illegal character strings, which (without context) generally do elicit more positive ERPs than legal words, but Laszlo and Federmeier have showed that illegal nonwords can in fact elicit N400 effects, and attribute the reduced N400 to factors such as lower (orthographic) neighbourhood competition \cite{laszlo_n400_2011,laszlo_wont_2012}. 

In addition to the N400 elicited by these familiar faces, the presentation of faces in comparison to other non-face stimuli (e.g., objects) elicits a negativity at 170ms (the N170) \cite{eimer_event-related_2000}. 

In their famous-faces speller, Kaufmann et al. \cite{kaufmann_flashing_2011,kaufmann_face_2013} demonstrate how these responses to (familiar) faces can be exploited by replacing the flashing of a letter in the matrix speller, by the superposition of a face over the letter (for each letter in a given row or column). This is shown to improve speller performance, with accuracy for 5 sequences of highlights at \textasciitilde{}95\% for the classic speller and \textasciitilde{}100\% for the famous faces in Kaufmann et al. (2011) \cite{kaufmann_flashing_2011}. In Kaufmann et al. (2013) \cite{kaufmann_face_2013}, both patients and healthy subjects were tested. There, again for 5 sequences, healthy subjects achieved \textasciitilde{}90\% and \textasciitilde{}99\% for the classic and faces speller respectively. The improvement for the patients was even more pronounced, improving from \textasciitilde{}77\% to \textasciitilde{}100\%, from the classic to the faces condition. In this followup-study they tested two additional face conditions: personal familiar faces and unfamiliar faces. The N400(f) effect for all three conditions were indistinguishable, while all significantly more negative than in the classic speller. The authors speculate that the presence of an N400 for the unfamiliar faces may have been due to an induced familiarity from the repeated presentations, however if unfamiliar faces are indeed analogous to (illegal) nonwords, it is not surprising that they elicit N400 activity \cite{laszlo_wont_2012}. 

In these studies a single face is used for all highlights across a session. Given the attenuating effect of repetition on the N400 amplitude (also applicable to faces \cite{bentin_effects_1994,eimer_event-related_2000}), this may not be ideal. While the increased performance of the face speller compared to the classic version shows that even a potentially attenuated effect is sufficient to increase the signal(s)-to-noise ratio of the speller, a collection of faces rather than a single face may improve performance further. Jin et al. \cite{jin_erp-based_2014} investigate this possibility, comparing a single face versus a multi-face paradigm. They find an increase in classification accuracy for the multi-face paradigm, (also specifically in the N400 window: 450-600ms). It is relevant to note that in their ERP grand averages, the single face condition had a consistently more negative amplitude than the multi-face condition in this time-window, which we would not expect if the N400 had attenuated due to a larger repetition effect in the single-face condition. Furthermore, their multi-face condition consisted of 5 faces; given that N400 repetition effects (for word stimuli) were still distinguishable at an interval of 20 intervening words before the repetition in previous literature \cite{rugg_event-related_1989}, a set of 5 may not be sufficient to measure the effect of this on the N400 (20-100 may be more appropriate). 

Overall, the faces speller paradigm offers a nice illustration of how the signal-to-noise ratio of an existing application can be boosted by adding an additional signal, and how automatic processing of meaning, as indexed by the N400, can be exploited for this purpose.

\subsection{Language processing detection for Disorders of Consciousness}

Another field where the N400 has been used is the detection of language processing in patients with Disorders of Consciousness (DoC). Such disorders range from coma, to unresponsive wakefulness syndrome (UWS; formerly referred to as vegetative state), in which patients appear awake but are completely unresponsive, to minimally conscious state (MCS), in which there is some evidence of behavioural responses to commands, but no communication is achieved \cite{laureys_unresponsive_2010,giacino_minimally_2002}. In these cases the lack of behavioural communication is attributed to impaired consciousness, as opposed to a disability arising from paralysis (e.g. complete locked-in state, CLIS). However, patients in the latter category may also have impaired levels of consciousness \cite{kubler_braincomputer_2007,kubler_braincomputer_2008}, so in practice the dichotomy may not be clear cut. 

Detection of language processing in patients with DoC can give insight into what level of cognitive processing is intact in a given individual. This is not a BCI application in the traditional sense of offering control to the user, but a BCI as a (offline) diagnostics or prognostics tool. Possibilities for providing communication BCIs for these patients have also been explored, with a number of studies showing that for certain patients, command following (i.e., performing a mental task when prompted) can be detected (e.g., in fMRI \cite{owen_detecting_2006,monti_willful_2010} or EEG \cite{cruse_bedside_2011,wang_enhancing_2017}).

Generally, studies that use EEG to determine cognitive processing in DoC patients, use a range of ERP tasks. These are most commonly oddball tasks, to elicit signals such as the MMN or the P300, and semantic tasks to elicit a N400 and/or a late positive component. Stimuli are presented auditorily, as hearing is assumed to be intact. In Kotchoubey et al. \cite{kotchoubey_information_2005}, for instance, a large group of patients (n = 78) with DoC are investigated, with three oddball and three semantic tasks. The percentage of patients for whom a significant ERP could be detected ranged from 8\% to 95\%, depending on the group (UWS or MSC) and the brain signal under consideration (N1/P2, MMN, P300, P600 or N400). In particular, \textasciitilde{}14\% (UWS) to \textasciitilde{}23\% (MCS) of patients a N400 could be detected for a given semantic task. These were similar to the percentages for a group of severely brain damaged but conscious control group. 

Such studies generally check whether the patients improve, and/or regain some degree of communication at a later date, to assess whether detection of these ERPs is predictive of patient outcome. In Kotchoubey et al. (2005) \cite{kotchoubey_information_2005}, patients that exhibited a MMN improved significantly more often than those that did not. For the N400, this effect was not significant (p=0.079), but in a more recent study by Steppacher et al. \cite{steppacher_n400_2013}, a significant correlation was found for the N400 and recovery. In that study, the P300, did not have a significant relationship with recovery, even though overall it was detected more frequently in patients,

Whether someone exhibits a given ERP can be decided in multiple ways: by performing one of various statistical tests to single trial data, or employing human raters to judge average ERPs. In all approaches there is some balance between the possibility of a false positive or false negative: judging someone to exhibit an ERP when in fact they do not, or vice versa \cite{kotchoubey_towards_2013}. This is still separate from the issue of whether or not someone who does not (appear to) exhibit an identifiable N400, is in fact processing the stimuli. The latter cannot be excluded as a possibility, as even in the healthy population, the N400 can not always be detected in single subjects, even with behavioural confirmation that they are processing the stimulus's content.

The degree to which this is a relevant factor depends on the rate in which the N400 can be detected in healthy subjects. While the N400 literature is large, this is not a question that is usually addressed in general neuroscience. In table~\ref{tab:population_overview} we have compiled a number of studies who do report individual detection rates, for both the general population and data from DoC studies. Studies who had a control group of healthy students are listed twice in the table, in split by their respective categories. Two studies employed a control group of patients with brain damage, but without a disorder of consciousness, as a closer control. These have been included with the label CTRL. We specify a number of variables that may affect the identification rate: the task and stimuli used, the number of stimuli collected per class, and the test used to determine significance. Both the sample of studies and the sample sizes within studies are small, limiting our ability to draw conclusions. 

\begin{figure*}[t]
\centering
\includegraphics[width=\textwidth]{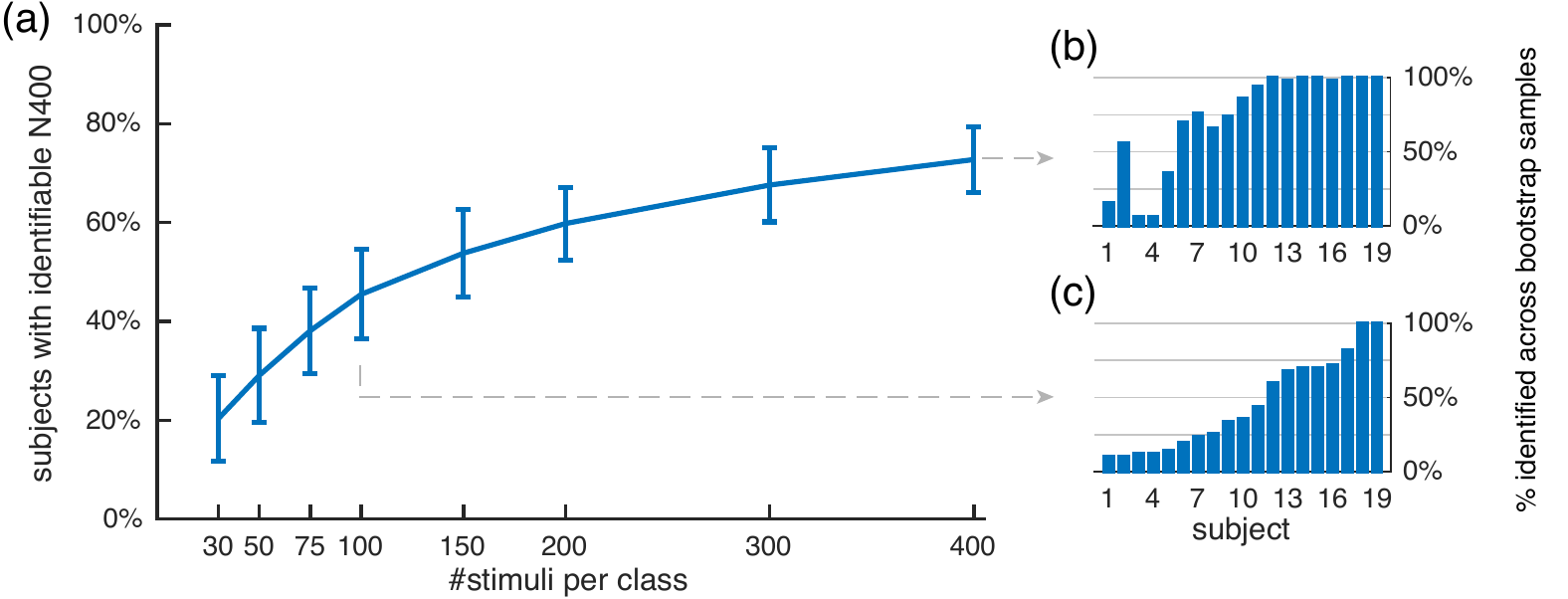}
\caption{\textbf{(a)} percentage of individuals (n=19) for whom a significant N400 can be detected for an increasing number of stimuli per class (data from Dijkstra et al. \cite{dijkstra_semantic_2018}). The plot represents the mean and standard deviations across 50 bootstrapped samples of each participant's data. \textbf{(b)} and \textbf{(c)}, per participant, the percentage of bootstrap samples in which an N400 effect could be identified using \textbf{(b)} 400 or \textbf{(c)} 100 (c) stimuli per class, sorted based on the 100-per-class detection percentage.}
\label{fig:N400_effects_pop}
\end{figure*}

In some sense the detection of a significant N400 effect in a single subject is an easier problem than making single trial predictions for BCI control. Table~\ref{tab:population_overview} shows that even this detection is/can be non trivial for the N400, even for healthy subjects. Furthermore, the data from the DoC studies suggest that in this application often relatively few stimuli per class are used to make these decisions. This may be due to time-constraints, as the N400 is often a single task in a task battery (i.e., combined with other tasks such as oddball paradigms), but likely hampers detectability. For instance, if we limit the number of trials per class to 50, in the Dijkstra et al. study \cite{dijkstra_semantic_2018} (listed in the table), then only about 6/19 individuals would have been identified as exhibiting an N400, rather than the 15/19 from using 400 brain responses per class. We determined this by running a cluster-based permutation test on the data from each individual subject, drawing a sample of trials per class (with replacement; 50 samples) from the full data of the respective individual, for a number of subset sizes ([30, 50, 75, 100, 150, 200, 300, 400]). The results of this analysis can be found in figure~\ref{fig:N400_effects_pop}.

At 100 stimuli per class only about 50\% of participants are identified to exhibit an N400, compared to when 400 stimuli per class are considered. The task employed in this study may not be ideal for detecting a N400: an implicit priming task was used in which the subject was given a target word to remember, followed by 1-10 probe words that they needed to assess the relatedness status to the target to. Yet, this analysis clearly illustrates that if the goal is to determine whether a given subject exhibits an N400, using 50-100 stimuli may produce a considerable amount of false negatives.

The fact that increasing the number of stimuli per class has such a effect on detection rates suggests that this is a signal-to-noise problem. Sculthorpe-Petley et al. \cite{sculthorpe-petley_rapid_2015}, approach this problem, by accruing information across participants, rather than increasing the number of presented stimuli. They train a Supoort Vector Machine on the averaged ERPs of each single subject, one for related and one for unrelated stimuli, and obtain a 92\% accuracy in a leave-one-subject-out training approach. We note however, that this appears to be the accuracy for predictions of both the related and unrelated ERPs for each subject, while the decision of whether a subject exhibits an N400 when only one of the two was classified correctly is non-straightforward. An open question is furthermore, whether this approach can be extended to the detection of the N400 in DoC patients, as they do not always exhibit a typical ERP \cite{kotchoubey_information_2005}.

If more time were available for these diagnostic tasks, there may be additional opportunities for assessing language processing. For instance, a subject that is listening to a story, may exhibit N400 effects based on the lexical properties of each given item, but not be tracking sentence level or discourse level content. Currently the paradigms outlined in table ~\ref{tab:population_overview}, only asses the mid level semantic content (primed words or [in]congruent sentences). The low level stimulus properties may help determine whether for a given individual an N400 can be detected, and if so, if this person also exhibit higher level language processing. This does assume that such lexical effects are easier (or as easy) to identify, which ought to be determined empirically in healthy subjects first. Similarly, manipulating discourse level semantics may allow to determine whether a patient accrues semantic context on longer timescales, though such manipulations may require longer sessions in order to collect sufficient data.

\newcommand{\mrl}[2]{
\begin{tabular}[c]
{@{} m{ #1 } @{} } 
#2
\end{tabular}}

\newcommand{\mrc}[2]{
\begin{tabular}[c]
{@{} >{\centering \arraybackslash} m{ #1 } @{} } 
#2
\end{tabular}}

\newcommand{\colsizeV}{0.175cm}
\newcommand{\colsizeP}{3cm}
\newcommand{\colsizeN}{1.5cm}
\newcommand{\colsizeT}{0.8cm}
\newcommand{\colsizeI}{1.4cm}
\newcommand{\colsizeNum}{1.2cm}
\newcommand{\colsizeD}{1.5cm}
\newcommand{\colsizeS}{4.5cm}

\begin{table*}
\footnotesize
\caption{
\label{tab:population_overview}Studies reporting detection rates of the N400 in individuals, from the general population or in patients with Disorders of Consciousness. }

\hspace*{-0.25in} 
\begin{tabular}{@{}m{\colsizeV}
m{\colsizeP}
>{\centering \arraybackslash}m{\colsizeN}
m{\colsizeT}
m{\colsizeI}
>{\centering \arraybackslash}
m{\colsizeNum}
>{\centering \arraybackslash}
m{\colsizeD}
m{\colsizeS}@{}}
\br

& paper & n & \hspace{-1ex}task &\hspace{-2ex}instruction \hspace{2ex}& \#stim /class & detection rate (\%) & statistic \\ \mr
\parbox[t]{2mm}{\multirow{4}[50]{*}{\rotatebox[origin=c]{90}{general population}}}{} & Bostanov \etal\cite{bostanov_t-cwt:_2006} & 36 & WP & passive &  200 & \mrc{1.3cm}{67 \\  94 } & \begin{tabular}[c]{@{}l@{}}t-CWT randomiziation test\textsuperscript{c}\\ t-CWT Hotelling\textsuperscript{c}\end{tabular} \\ \addlinespace

 & Daltrozzo \etal \cite{daltrozzo_cortical_2009} & 20 & \begin{tabular}[c]{@{}l@{}}WP\\ SC\end{tabular} & passive & \begin{tabular}[c]{@{}l@{}}60 \\ 50\end{tabular} & \begin{tabular}[c]{@{}l@{}} 45 \\ 60 \end{tabular} & t-CWT Hotelling\textsuperscript{c}\\ \addlinespace
 
 & \multirow{4}{*}{Cruse \etal\cite{cruse_reliability_2014}} & \mrc{\colsizeN}{12 \\ 12 \\ 12} & \mrc{\colsizeT}{ \\ WP\textsuperscript{a} \\ \\} & \mrl{\colsizeI}{overt\\ covert\\ passive} & \mrc{\colsizeT}{ \\ 100 \\ \\} & \mrc{\colsizeD}{ 75  \\  58  \\ 0 } & \multirow{4}{\colsizeS}{cluster permutation test\textsuperscript{d}} \\
 &  & 12 & WP\textsuperscript{b} & passive & 100 &  50  &  \\
 &  & 12 & SC & passive & 100 & 17  &  \\ \addlinespace
 
 & Sculthorpe-Petley \etal\cite{sculthorpe-petley_rapid_2015} & 100 & SC & passive & 30 & \begin{tabular}[c]{@{}l@{}}26 \\ 92 \end{tabular} & \begin{tabular}[c]{@{}m{\colsizeS}@{}} cluster permutation test\textsuperscript{d}\\ cross-subject SVM\textsuperscript{e} \end{tabular} \\ \addlinespace
 
 & Rohaut \etal\cite{rohaut_probing_2015}& 19 & WP & passive & 68 & 42  & t-test criterion\textsuperscript{f} \\ \addlinespace
 
 & Geuze \etal\cite{geuze_detecting_2013} & 12 & WP & covert & 200 & 100  & binomial confidence interval\textsuperscript{g} \\ \addlinespace
 
 & Dijkstra \etal\cite{dijkstra_semantic_2018} & 19 & WP & covert & {\scriptsize $\sim$}500 &  \mrc{\colsizeD}{ 63 \\ 79 } & \mrl{\colsizeS}{binomial confidence interval\textsuperscript{g} \\ cluster permutation test\textsuperscript{d}}\\ 
 
 \mr 
 
 \parbox[t]{2mm}{\multirow{17}{*}{\rotatebox[origin=c]{90}{Disorders of Consciousness}}}{} 
 & Schoenle \etal\cite{schoenle_how_2004} & \mrc{\colsizeN}{43 \textsc{uws}\\23 \textsc{nevs} \\ 54 \textsc{ctrl}} & SC & passive & 100 & \mrc{\colsizeD}{39  \\ 77  \\ 90}& human raters (experts)\textsuperscript{h} \\ \addlinespace
 
  & Kotchoubey \etal\cite{kotchoubey_information_2005} &  \multicolumn{5}{l}{ \begin{tabular}[c]{@{}>{\centering \arraybackslash}m{\colsizeN}m{\colsizeT}m{\colsizeI}>{\centering \arraybackslash}m{\colsizeNum}m{\colsizeD}@{}} 38 \textsc{uws}  & \mrl{\colsizeT}{WP \\ SC} & \multirow{5}{\colsizeI}{passive} & \multirow{5}{*}{50} & \mrc{\colsizeD}{14  \\ 23 } \\ 
  38 \textsc{mcs} & \mrl{\colsizeT}{WP \\ SC} &  & & \mrc{\colsizeD}{20  \\ 18 } \\
  10 \textsc{ctrl} & \mrl{\colsizeT}{WP \\ SC} &  & & \mrc{\colsizeD}{22  \\ 14 }\\
 \end{tabular}} & one-tailed ANOVA\textsuperscript{i} \\ \addlinespace 
 
 & Daltrozzo \etal\cite{daltrozzo_cortical_2009} &  42 coma & \mrl{\colsizeN}{WP \\ SC} &  passive & \mrc{\colsizeNum}{60 \\ 50} & \mrc{\colsizeD}{ 17  \\  7 }  & t-CWT Hotelling\textsuperscript{c} \\ \addlinespace
 
 & Steppacher \etal\cite{steppacher_n400_2013} & \multicolumn{5}{c}{ \begin{tabular}[c]{@{}>{\centering \arraybackslash}m{\colsizeN}m{\colsizeT}m{\colsizeI}>{\centering \arraybackslash}m{\colsizeNum}>{\centering \arraybackslash}m{\colsizeD}@{}} 53 \textsc{uws}  & SC & covert & 100 & \mrc{\colsizeD}{ 32  \\  15 } \\ 
 39 \textsc{mcs} & SC & covert & 100 & \mrc{\colsizeD}{ 44 \\  23 } \\
 \end{tabular}} & t-CWT\textsuperscript{c} \newline human raters (experts)\textsuperscript{h}\newline t-CWT\textsuperscript{c} \newline human raters (experts)\textsuperscript{h} \\ \addlinespace
 
  & Rohaut \etal\cite{rohaut_probing_2015} & \mrc{\colsizeN}{15 \textsc{uws} \\ 14 \textsc{mcs}} & WP & passive & 68 & \mrc{\colsizeD}{  7  \\ 36 } & t-test criterion\textsuperscript{f} \\ 
 
\br

\multicolumn{8}{p{1.1\textwidth}}{\scriptsize WP: word priming task |~SC: sentence congruence task |~UWS: unresponsive wakefulness state |~MCS: minimally conscious state |~NEVS: near vegetative state (approx. between uws and mcs) |~CTRL: control group of brain injured, but conscious patients} \\

\scriptsize
\begin{tabular}[c]{p{0.5\textwidth}p{0.5\textwidth}}
\noindent\textsuperscript{a} relatedness based on shared semantic features \cite{mcrae_semantic_2005}

\noindent\textsuperscript{b} relatedness based on association norms \cite{nelson_university_2004}

\noindent\textsuperscript{c} Student's t-statistics applied to a Continuous Wavelet Transform \cite{bostanov_t-cwt:_2006}

\noindent\textsuperscript{d} non-parametric cluster-based permutation test \cite{maris_nonparametric_2007} 

\noindent\textsuperscript{e} Support Vector Machine trained on the average ERPs of n-1 subjects and tested on the n\textsuperscript{th} subject
& 
\noindent \textsuperscript{f} unpaired t-test criterion: p $\leq$ 0.05 on a minimum of 5 samples and 10 electrodes

\noindent \textsuperscript{g} a binomial confidence interval used to distinguish prediction accuracies from chance

\noindent \textsuperscript{h} averaged ERPs evaluated by neurophysiologists.

\noindent \textsuperscript{i} one-tailed ANOVA of factor condition (related vs unrelated) on (windowed) ERP amplitude\\
\end{tabular}

\end{tabular}
\normalsize

\end{table*}

\subsection{Inferring information about the user's active mental context}

The fact that the active mental context of a subject modulates the N400 to a given stimulus can be exploited for BCI purposes by presenting stimuli of which the semantic content is known and, through presentation of stimuli with varying semantic content and decoding the (absence) of relationships, inferring this mental context. 

One line of research in this direction is that of relevance detection: figuring out what a user is searching for based on their brain response to specific words or images. For instance, Wenzel et al. 2014 \cite{wenzel_real-time_2017} attempt to decode a category of interest based on brain responses to stimuli words out of various categories. To make the setting more natural, multiple words were presented simultaneously, distributed spatially across the screen. The time-lock of the brain response to the stimulus was achieved with an eyetracker, based on the moments of eye fixation. In the (online) test phase of the experiment, subjects picked a single category out of 5 as a target and attended stimuli of all 5 categories. Attended stimuli disappeared (based on gaze detection), and were replaced with new stimuli. Incoming data was then analysed to update ranking of the expected target category, online. After a 100 stimuli a trial ended and new categories were selected. 

Given the semantic nature of this task, this should elicit a N400 response, and the reported results indeed show a N400-like negativity in Cz for stimuli that were not members of the target category. It is important to note that the relevant-irrelevant difference was markedly reduced in the online phase. The authors suggest this may be due to a change in task (during training subjects were asked to count, during testing only to search for relevant terms). However, since the experiment used a total of 17 categories with 20 stimuli each, subjects would have seen most stimuli at least once during training, and while during a \textit{given} trial in the testing phase subjects saw each stimuli only once, categories would return in other trials, leading to repetitions of stimuli. It is therefore possible this attenuation can be attributed to a N400 repetition effect. In addition to the N400, they found a positive component in left posterior electrodes (e.g., P9), from 200-600 ms. This positive component was more pronounced in the online phase, and persisted past 800 ms. Furthermore, in this online phase, this positive component was more widespread, showing a left lateralised effect, that showed up as a late positive component (following the attenuated N400), in electrode Cz. Performance for this BCI was measured based on the ranking of the correct category out of all five categories, with a mean rank of 1.68 (chance level is 3) across all subjects (range 1.12-2.47). 

In Golenia et al. \cite{golenia_implicit_2018} a similar paradigm is used to detect which concept is relevant to a users search, but here the aim is to disambiguate between multiple meanings of a term using images. Fixation-related potentials were collected in response to pictures each depicting one of two interpretations of a search term (e.g., \textit{bass} the fish, or \textit{bass} the instrument). While this task can be interpreted as presenting related and unrelated stimuli in context of a target (the intended meaning), this task did not appear to elicit a N400 effect. Instead, there was a late effect (>500ms), more negative for images with the intended meaning (i.e., related), compared to the images with the alternative meaning, that could successfully be exploited for classification. Picture stimuli have been shown to elicit N400 effects in priming studies \cite{barrett_event-related_1990}, so an N400 could have reasonably been expected and we have no clear hypothesis for why this paradigm would not elicit one.

In our own research group we have pursued a similar application, presenting stimuli to deduce information about the active mental context, but aimed for word selection, rather than category selection or word/concept disambiguation. Our studies thus far have been offline experiments, that aim to determine the suitability of the N400 for the intended task. Specifically, a first study determined whether or the N400 could be decoded in single trials by determining whether a given stimulus (i.e., the probe) was related to the active mental context (the immediately preceding prime) \cite{geuze_detecting_2013}. These stimuli were presented in word-pairs where half of the primes were related and half unrelated. Across 12 subjects, classification accuracies ranged from 54\% to 67\%, (average \textasciitilde{}60). 

In a follow-up study we aimed to determine whether or not the N400 can reliably be detected when the prime word is not presented, but actively recalled by the subject, while multiple consecutive probe words are presented. An earlier study by Van Vliet et al. had already established that it is not necessary to actually present a prime word, it is sufficient for the subject to actively recall it \cite{van_vliet_guessing_2010}. In this consecutive probing experiment a trial started by supplying the subject with a target word to remember, after which one to ten probe words were presented, with probe words either strongly related or unrelated to the target. No attenuation between the first and 9$^th$ or 10$^th$ probe was detected, suggesting that multiple consecutive probes can be used to elicit information about a target concept on the users mind. At the same time, classification rates were again low: accuracies for single probes ranged from 50\% (indistinguishable from chance level) to 72\% (mean 58\%), and for approximately 1/3 of subject the accuracy was not distinguishable from the (bonferroni-corrected 95\%) confidence interval of chance performance. While it is not an unknown problem for a subset of users unable to use a BCI \cite{allison_could_2010}, this can be a concern for BCI development. 

Low single trial classification rates can be overcome by accumulating data over multiple brain responses. Given the sensitivity of the N400 to repetition of stimuli, this does not have to be limited to aggregation of information over a single repeated probe, but can consist of accumulating information across different probes. The Wenzel et al. relevance detection study is an clear example where category detection is possible, when the set of categories is small, and inference is performed on the aggregating over a large number of stimuli presentations (\textasciitilde{}100 stimuli). Their fixation-related-potentials approach allows users to decide their own pace, and the short fixation times (\textasciitilde{}200 to \textasciitilde{}300ms), show that requiring many stimuli is not necessarily a problem, when single responses can be obtained quickly. However, it remains a question how well this approach can be expanded to allow selections from a larger number of categories or concepts. In the Dijkstra et al. study, no online phase was included, but simulations were used to estimate how well the correct target could be inferred from the others (\textasciitilde{}120 possible targets). The results from these simulations indicated that only for about 5 subjects, 100 stimuli would be sufficient to have the correct target in the top 3 guesses (on average; see Dijkstra et al. for more details \cite{dijkstra_semantic_2018}). Such simulations make various (implicit) assumptions and have limited interpretability compared to results from an online test, but this illustrates that extending this approach to larger concept dictionaries (e.g., 1000), would likely require an impractical number of consecutive probes, with this approach.

The limiting factor here is likely the signal-to-noise ratio of the N400 that, for the average subject, appears to be lower than that of e.g., the P300, given the difficulty of even finding a significant effect in a given individual (see figure~\ref{fig:N400_effects_pop}). A possible solution may be to increase the signal-to-noise by evoking an additional signal, e.g., by coupling the probing approach with a task to elicit a P300, similar to the use of faces in the famous faces speller. This approach was used in Geuze et al. (2014) \cite{geuze_towards_2014}, in which subjects were instructed to press a button for all related stimuli in a stream, intended to elicit both a P300 and an event-related desynchronisation (ERD) over the motor cortex. The authors found a positive ERP for related stimuli from 300ms to 1 second, and obtain classification rates ranging from 59\% to 77\% using time and time-frequency features. These classification rates may depend on the P300, the ERD, and/or the N400, and it is unclear to which degree each contribute. Regardless, they are higher than the 50\%-72\% for cross-validated, offline classification of the N400 in a similar task \cite{dijkstra_semantic_2018}. The Wenzel study, notably, also established an ERP in addition to the N400, a left-lateralized late positivity for \textit{un}related stimuli. This matches the LPC or P600 responses that are sometimes reported to co-occur with N400 tasks, and that have been interpreted as late P300's by some \cite{leckey_p3b_nodate}. However, this response is in the opposite direction as ERP in the Geuze et al. (2014) study, where the unrelated stimuli were more negative. Furthermore, in the Wenzel study, the relative strength of the N400 and this P600 differed between the training and test phases (where subjects received an explicit task, or no task, respectively). Further research would be required to understand the different factors at play here. While adding deliberate responses such as an ERD or a P300 to this approach may certainly be successful, it does detract from one of the main draws of the N400, its more automatic nature: an intuitive task ("process these stimuli") is likely less cognitively taxing than an deliberate task ("do \textit{X} when this stimulus appears").

These probing paradigms can also be improved by determining the right stimuli to present at a given time, similar to the optimisation of flashing patterns in P300 spellers. By updating after each new stimulus, the information accumulated thus far can then be exploited to select which stimulus would be particularly informative or to determine whether sufficient confidence has been reached to make a prediction. Which stimulus to present next, in this scenario, is analogous to asking for which instance (i.e., stimulus) the system would prefer to receive a label (i.e., the relatedness status) next. This is a question explored in the field of active learning, and techniques from this field may be easy to adopt (see Settles, 2012, for an overview \cite{settles_active_2012}). A simple strategy is, for instance, to present the stimulus for which the distribution across predicted/expected response(s) currently has the highest variance (known as `uncertainty sampling'). More sophisticated strategies avoid potential pitfalls of uncertainty sampling, but can become computationally expensive \cite{settles_active_2012}.

\section{Discussion}
We have outlined how the N400 can be elicited in various ways (visually or auditorily, with linguistic or other meaningful stimuli) and how its amplitude is sensitive to a range of experimental manipulations and parameters (active or passive mental context, stimulus properties). 
Analyses of BCI studies across three application areas suggests that the N400 can successfully be exploited for BCI purposes, but that the signal-to-noise ratio is a limiting factor, with signal strength also varying strongly across subjects.
We have shown how this offers a range of opportunities for exploitation, evidenced by the applications discussed, as well as how it poses difficulty in designing BCI paradigms, due to both an apparent low signal-to-noise ratio and a high potential of confounding variables.

One unaddressed aspect is how to obtain labelled stimuli for eliciting these effects. For linguistic stimuli there many resources. For instance, for determining the relatedness of two words there are association norms databases (e.g., \cite{kiss_associative_1973,deyne_word_2008}). Alternative measures of relatedness (e.g., similarity) can be extracted from WordNet \cite{fellbaum_wordnet_2012} (see e.g., \cite{budanitsky_evaluating_2006}). Word embeddings from Computational Linguistic represent the meaning of a word as a (high dimensional) vector, that can be used to compute relatedness (e.g., word2vec \cite{mikolov_distributed_2013,noauthor_word2vec_2013} or LSA \cite{landauer_introduction_1998,noauthor_lsa_nodate}).
Lexical properties (e.g, word frequency and neighbourhood measures) can be found in lexical databases (e.g., \cite{balota_english_2007,baayen1995celex}). Databases also exist for sentence stimuli (e.g., \cite{block_cloze_2010}), and word-embeddings can be extended to represent larger contexts by averaging over terms (see e.g., \cite{broderick_electrophysiological_2018}). For non-linguistic stimuli it is less trivial to obtain labelled stimuli. For pictures Golenia et al. used the Flickr API that allows images to be retrieved based on tags \cite{golenia_implicit_2018}. For face images in particular, there is a Microsoft database for celebrity faces \cite{noauthor_msra-cfw_nodate}, though it may be non-trival to make these suitable for insertion into e.g., a matrix speller.

We have discussed three lines of current research that use a N400 for BCI purposes: (1) enhancing the matrix speller by using the (N170 and) N400 response to familiar faces, (2) the detection of language processing in patients with Disorders of Consciousness (DoC), and (3) the use of probe stimuli to infer information about a user's mental context. The success of the use of famous faces in matrix spelling applications in boosting spelling performance shows that the N400 can successfully be exploited for BCI purposes. It would be of interest to see whether this performance boost extends to covert speller paradigms, where the differences between face presentations for attended and unattended letters derive only from the users interpretation of the relevance of the stimulus, not the location of its presentation (foveal or peripheral) \cite{treder_covert_2010}. Furthermore, the fact that repetitions of (face) stimuli have been established to attenuate the N400 amplitude provides an opportunity to explore whether presenting larger sets of faces may boost performance further.

The diagnosis of language processing in patients with DoC on the basis of the N400, was shown to have prognostic value: patients that can be shown to exhibit an N400 are more likely to improve in their condition or recover. At the same time it is clear from comparative paradigms in healthy controls that the detection of the N400 even in the general population is non-trivial. An analysis in which we increased the number of stimuli per class that are considered when making the decision for a given subject, illustrated that detection rates continue to rise well beyond the number of stimuli typically considered when judging the presence or absence of an N400. Low detection rates of the N400, when considering small numbers of trials, can thus easily be attributed to low signal-to-noise ratios as opposed to an indication of a lack of linguistic processing. Next to increasing the number of stimuli presented to the subject in order to make a judgement, this low signal-to-noise ratio can also be overcome by accruing data across individuals, as demonstrated by Sculthorpe-Petley et al. \cite{sculthorpe-petley_rapid_2015}. However whether ERPs from DoC patients are similar enough for this to work remains an open question. To summarise, it is clear that when the N400 is used as a proxy for determining language processing in an individual, the potential of a false negative (deciding a patient does not process language when in fact they do), is substantial, regardless of the statistical method, evidenced by the non-perfect detection rates in healthy subjects. The probability of a false positive, conversely, will depend mostly on the approach. Statistical approaches, if properly applied, are typically designed to have a 5\% probability of a type I error (i.e., a false positive). Using human raters gives a less clear control of the false positive rate, though the results from Steppacher et al. \cite{steppacher_n400_2013} show that it is not necessarily the case that human raters are more likely to produce positive judgements, than a statistical method (see table~\ref{tab:population_overview}).

Patients with a disorder of consciousness, whom have been determined to exhibit an N400, could also be a target population for BCI paradigms that attempt to infer information about the active mental context of the user, using the probing approaches. Whether this has a potential for success depends on whether BCI control beyond command following is even theoretically possible for DoC patients, and may depend on the specific diagnosis (e.g., UWS or MCS).

This application of inferring information about the active mental context of the user was the third application we discussed. On one hand there is evidence that this approach can be used successfully to determine which of a small number of categories a subject aims to select ('has on mind') \cite{wenzel_real-time_2017}. By using fixation-related potentials, stimuli can be presented to a user at a fast speed, as the inter-stimulus interval is determined by the subject, through their gaze. On the other hand, approaches where this is attempted to infer concepts from a larger space, the low signal-to-noise ratio becomes a problem, and subject variability is amplified. Here limits to the exploitation of the N400 become clear: Single trial classification rates are low (\textasciitilde{50-75\%}), and there is high variability across subjects in the strength of the N400 (see also figure~\ref{fig:N400_effects_pop}), if it can be identified at all. There are possibilities for improving these paradigms: e.g., by inducing an additional signal such as the P300 or by using techniques to do informed probe selection, however, it is unclear whether this could ever lead to an application where subjects can select from a large concept space (e.g, >1000).

The BCI applications explored so far focus on sentence level and priming effects. However, as higher level discourse or context also affect the N400, it may be possible to design BCI paradigms that exploit such effects. For this it is useful to have representations of meaning coherence at a higher level. As alluded to earlier, word embedding vectors can used for this purpose. For instance, 
Broderick et al. \cite{broderick_electrophysiological_2018}, demonstrate how a regression to a word-embedding measure of relatedness can be used to measure comprehension of an audiobook. The authors fit Temporal Response Functions (TRFs) of EEG responses to word (dis)similarity values extracted from word2vec, using averaged word vectors to represent sentence level context. These TRFs subsequently exhibited a stronger negativity in the N400 time-range when fitted to an attended story, compared to a, simultaneously presented, unattended story. While in that study the authors use the embeddings to calculate sentence level congruence, this measure could also be extended to asses higher level semantic content, by averaging word vectors over larger units of text (e.g., paragraphs and sections). 

Combining measures of comprehension at different levels, could allow indexation of the level of understanding of a given text. Such an approach could be used to detect text comprehension, which may be of interest for a DoC language processing application, but also for detecting whether someone is able to understand a given text in a non-patient population. It may be non-trivial to develop an application that detects the level of expertise in a academic field (do word-embeddings then need to be trained on this academic literature?), but a more straightforward application may be to determine appropriate reading levels in children. A study by Holcomb et al. (1992) \cite{holcomb_visual_1992} investigated the N400 in ages between 5 and 26 and found N400 amplitudes to be larger for younger subjects, with amplitudes stabilising around age 15-16. A comprehension detection application may thus prove especially sensitive when aimed at children. Such an approach could also be applied to adult second language learners. These learners have, for instance, been shown to have distinguishable N400 responses to foreign language words and pseudo-words after only 14 hours of instruction in the respective language \cite{mclaughlin_neural_2004}. On a more speculative note, N400-like waves have even been identified during sleep \cite{ibanez_erps_2009}, which may allow for evaluation paradigms for which subjects need not be awake.

These applications do not only exploit the active contextual effects ("are you processing the meaning of this text"), but also whether the passive requisite knowledge is present for understanding. It is good to point out that this can also be used to elicit more personally identifying information about an individual. A strong illustration of this is a recent study in which congruency effects were elicited using Harry Potter related stimuli, and the authors found a correlation between the N400 strength to these stimuli and the self-reported familiarity with the Harry Potter fandom \cite{troyer_harry_2018}. Such an approach could be extended to predict an individual's familiarity with the books, raising ethical considerations: an application based on \textit{active} mental context, can be subverted more easily than an application that probes your \textit{passive} (world-) knowledge. While (un)familiarity with Harry Potter may not constitute sensitive personal information for most, these methods could perhaps be used to elicit other personal details, such as political or religious beliefs. If BCI applications become more widely adopted, it may become possible for malicious parties to covertly extract such information, by strategically inserting certain linguistic stimuli within another application.

Whether these suggested applications are viable as BCI paradigms remains an open question. It would be useful to have a better idea of the relative size of the different manipulations and a predictive model that can take the effect size or signal-to-noise ratio for a given manipulation and determine whether the N400 effect is sufficiently strong that it is suitable for an intended BCI task. However, it appears size of the N400 amplitude difference is highly dependent on the manipulation, task and subject characteristics, making such an estimation non-trivial.

Overall, the N400 offers access to someone's active mental context or their passive knowledge and associations, which makes it highly interesting for BCI applications, yet its relatively low signal-to-noise ratio makes it difficult to exploit. This makes the N400 most suitable for applications where longer integration times (i.e., more stimulus information that is averaged/accumulated over) are acceptable, such as the suggested use of determining language processing through presentation of texts or audiobooks.

\ack{We thank Mante Nieuwland and Ceci Verbaarschot for helpful comments.}

\newcommand{\newblock}{}
\bibliographystyle{unsrt}
\bibliography{sample}

\end{document}